# Laser wakefield acceleration with high-power, few-cycle mid-IR lasers

Daniel Papp[1], Jonathan C. Wood[2], Vincent Gruson[3], Mina Bionta[3], Jan-Niclas Gruse[2], Eric Cormier[4], Zulfikar Najmudin[2], François Légaré[3], Christos Kamperidis[1]

[1]: ELI-ALPS, ELI-HU Non-profit Ltd., Szeged, Hungary; [2]: The John Adams Institute for Accelerator Science, Blackett Laboratory, Imperial College London, UK; [3]: Institut National de la Recherche Scientifique, Centre Énergie, Matériaux, et Télécommunications, Varennes, Québec, Canada, [4]: CELIA, Université Bordeaux, Talence, France

**Abstract**

The study of laser wakefield electron acceleration (LWFA) using mid-IR laser drivers is a promising path for future laser driven electron accelerators, when compared to traditional near-IR laser drivers operating at 0.8 - 1 μm central wavelength ($\lambda_{laser}$), as the necessary vector potential ($a_0$) for electron injection can be achieved with smaller laser powers due to the linear dependence on $\lambda_{laser}$. In this work, we perform 2D PIC simulations on LWFA using few-cycle, high power (5-15 TW) laser systems with $\lambda_{laser}$ ranging from 0.88 – 10 μm. Such few-cycle systems are currently under development, aiming at Gas High Harmonics Generation applications, where the favourable $\lambda_{laser}^2$ scaling extends the range of the XUV photon energies. We keep $a_0$ and $n_e/n_{cr}$ ($n_e$ being the plasma density and $n_{cr}$ the critical density for each $\lambda_{laser}$) as common denominators in our simulations, allowing for comparisons between drivers with different $\lambda_{laser}$, with respect to the accelerated electron beam energy, charge and conversion efficiency. While the electron energies are mainly dominated by the plasma dynamics, the laser to electron beam energy conversion efficiency shows significant enhancement with longer wavelength laser drivers.

## I. Introduction

Laser plasma accelerators, proposed in 1979 [1], utilize the high electric field gradients (in the order of 100 GV/m) achievable in plasmas for accelerating electrons to GeV energies in distances up to a few cms [2-5], with self-focusing extending acceleration lengths as far as 100 Rayleigh ranges [6]. The short-pulse lasers required to excite these plasmas allowed the development of high brightness photon sources [7,8].

The ponderomotive force based scaling laws [9] describing the electron Laser Wakefield Acceleration (LWFA) process predicts some advantage of longer wavelength lasers, i.e. that wakefield acceleration can be achieved at lower laser powers than with current laser systems with central wavelength around 800 nm. LWFA experiments were proposed for multi-TW mid-IR laser systems under construction, at 2 μm wavelength in China [10], and at 10 μm at Argonne National Laboratories [11] to investigate this wavelength range. LWFA electron acceleration experiments in the self-modulated range were also conducted using a 3.9 μm laser, producing a thermal electron spectrum extending over 10 MeV [12].

One property of few-cycle laser systems is that with the shorter driving laser pulse, shorter plasma wavelengths $\lambda_p$ (and plasma bubble sizes) – and increased plasma electron densities – are possible without the laser pulse overlapping and interfering with the plasma bubble or the accelerated electron bunch. The larger electron density itself allows plasma self-focusing, self-guiding, and also facilitates the injection of electrons. E.g. a 880 nm, 5TW laser system with a longer, $\tau_l$ = 28 fs pulse duration and corresponding 140 mJ energy has a laser pulse length of 8.4 μm. If the plasma density is set that the driving pulse should not interfere with the electron bunch in the first bubble, as shown in Fig. 1, then the plasma wavelength must be at least double of the laser pulse length, i.e., $\lambda_p > 2 \times \tau_l$, which corresponds to a maximum plasma density of $n_e$ = 0.0027 $n_c$, where the critical plasma density is $n_c = \varepsilon_0 m_e \omega^2 / e^2$. At this relatively low plasma density plasma self-focusing will not occur (the nominal self-focusing treshold density for the aforementioned laser parameters is 0.0037 $n_c$), and is also too low for either self-injection or ionization injection. However, if the driving laser pulse is shorter at the same laser power, e.g. 35 mJ in 7 fs, the maximum electron density – with the conditions as above - can be as high as $n_e$ = 0.044 $n_c$ before the laser pulse would disrupt the acceleration process. This density permits both injection and self-focusing at the relatively modest power of 5 TW.

This reduced laser power requirement would allow LWFA to be driven by relatively low energy, few-mJ TW lasers that operate at higher repetition rates, producing upto 22 MeV electron bunches with 4 fs lasers at 10 Hz repetition rate [13]. The LWFA process was demonstrated at 1 kHz repetition rates in the blowout regime, producing a quasi-monoenergetic 5 MeV electron spectrum [14]. In earlier experiments 0.5-1kHz repetition-rate LWFA experiments were conducted at several laboratories, producing a self-modulated electron beams [15-17].

The SYLOS laser system developed for ELI-ALPS is the first multi-TW, 1 kHz repetition-rate few-cycle laser system, operating at 880 nm central wavelength, with 55 mJ laser pulses under 9 fs [18]. This laser system is under further development, to be upgraded to 35 mJ/7 fs ("SYLOS-2A"), and later to 100 mJ/5fs ("SYLOS-2B") pulse properties. This laser system is suitable for driving a 1 kHZ repetition-rate LWFA electron accerator beamline, providing 20+ MeV (projected) electrons with reasonable bunch charge.

This paper is investigating, via 2D simulations, the wavelength scaling of 5 and 15 TW few-cycle laser pulses in a parametric study, where the expected laser parameters of the ELI-ALPS SYLOS laser system were scaled for infrared wavelengths.

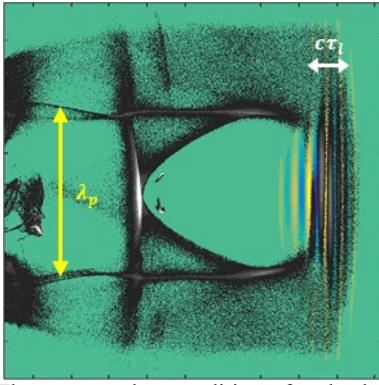

Fig. 1. The non-overlap conditions for the laser pulse length and plasma electron bubble size in the blowout regime – transverse electric field shown in blue and yellow

## II. PIC Simulations

The parametric study was conducted using the 2-D particle-in-cell code EPOCH2D [19]. The resolution of the simulation grid varied from $\lambda/33$ (at 880 nm) to $\lambda/160$ (at 10 μm) in the propagation direction, with 1:2 cell aspect ratio. The simulation domain varied from 50 μm×50 μm to 250 μm ×250 μm, with the frame moving with the laser pulse. The laser pulse was assumed to be Gaussian both spatially and temporarily. Laser pulse propagation was described by setting entrance boundary amplitude and phase according to the paraxial formalism. The plasma in the simulations was neutral He gas with ionization potentials of 24.6 eV and 54.4 eV. Three ionization processes available in the code (field, multiphoton and barrier suppression ionization) were taken into account. The gas density was assumed to have a flat profile with 100 μm ramps at both ends.

The electron bunch energy (per unit length) and average electron energy was determined from the electron $x$-$p_x$ phase-space distribution, while the laser pulse energy (per unit length) value was determined from the transverse electric field during the early stages of the simulation. The ratio of the electron bunch energy and laser pulse energy corresponds to the laser-to-electron-bunch energy conversion efficiency. From the efficiency, nominal laser pulse energy, and average electron energy, the bunch charge could be inferred.

## III. Simulation parameters and results

The wavelengths investigated in the study were chosen to be 880 nm (of the ELI-ALPS SYLOS laser system), 1.8 μm (representing several gain media around this wavelength), 3.2 μm (within the wavelength range of the ELI-ALPS Mid-IR laser system [20]) and 10 μm (of $CO_2$ lasers). The parameters kept constant were the laser pulse length of 2.4 laser cycles, the laser power of 5 TW, and the electron density of $n_e$ = 0.02 $n_c$ ($n_c$ being the plasma critical density), unless noted otherwise. For a second set of simulations, the laser power was taken to be 15 TW, similar to laser pulse properties that are expected to be available using the SYLOS-2B system (100 mJ 5 fs). The parameters and results of the simulations are shown in Table 1.

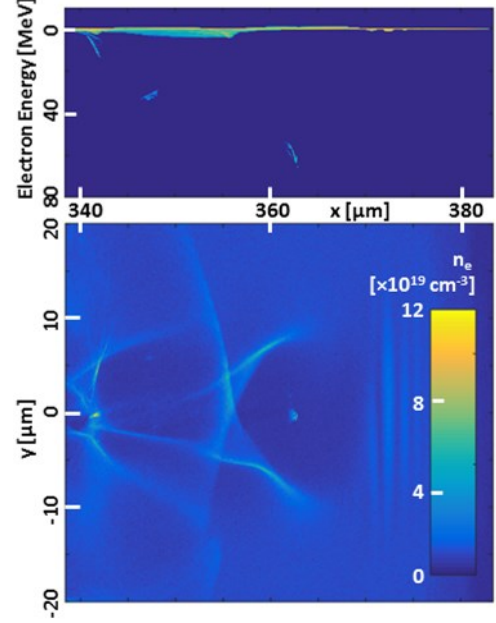

Fig: 2. A 2D electron density map (bottom) and $x$-$p_x$ phase-space plot (top, logarithmic, arb. u.) for the 5 TW, 1.8 μm case

For the 5 TW series, the 880 nm case has marginal electron injection in the second bubble. For all the longer wavelengths, a distinct quasi-monoenergetic electron bunch develops (Fig. 2). As the wavelength increases, the laser-to-electron conversion efficiency jumps from 0.43% (880 nm) to 4.2% (1.8 μm) to 9.9% (3.2 μm), which is also represented in high bunch charges. In contrast, both average and maximum electron energy decreases with wavelength. This decrease of electron energy is also present for the 15 TW cases. The dependence of the bunch charge/pulse energy ratio and the average bunch energy is shown in Fig. 3 for both the 5 TW and 15 TW cases, with the charge/pulse energy ratio increasing with the wavelength upto 3.2 μm then plateauing, and the average bunch energy having a maximum at 1.8 μm wavelength.

| λ (μm) | 0.88 | 1.8 | 3.2 | 10 | 0.88 | 1.8 | 3.2 | 10 | 3.2 |
|---|---|---|---|---|---|---|---|---|---|
| Laser Power (TW) | 5 | 5 | 5 | 5 | 15 | 15 | 15 | 15 | 15 |
| Laser energy (mJ) | 35 | 72 | 128 | 400 | 105 | 216 | 384 | 1200 | 384 |
| Duration (fs) | 7.0 | 14.4 | 25.6 | 80 | 7.0 | 14.4 | 25.6 | 80 | 25.6 |
| $w_{FWHM}$ (μm) | 6.4 | 13 | 23.1 | 72.2 | 6.4 | 13 | 23.1 | 72.2 | 23.1 |
| Intensity ($/10^{18}$ Wcm$^{-2}$) | 10 | 2.4 | 0.77 | 0.075 | 30 | 7.3 | 2.3 | 0.23 | 2.3 |
| $n_e/n_c$ | 0.02 | 0.02 | 0.02 | 0.02 | 0.02 | 0.02 | 0.02 | 0.02 | 0.015 |
| $n_e$ ($/10^{18}$ cm$^{-3}$) | 29 | 6.9 | 2.2 | 0.77 | 29 | 6.9 | 2.2 | 0.77 | 1.65 |
| $W_{bunch}$ (MeV) | 15 | 61 | 52 | 48 | 48 | 59 | 48 | 43 | 85 |
| $W_{max}$ (MeV) | 56 | 68 | 64 | 60 | 109 | 104 | 95 | 80 | 145 |
| Efficiency | 0.43% | 4.2% | 9.9% | 8.8% | 20% | 37% | 44% | 39% | 33% |
| Charge (pC) | 10 | 50 | 242 | 730 | 430 | 1300 | 3500 | 10900 | 1500 |

Table 1: The sets of laser/plasma characteristics used in the parametric study, and resulting electron bunch parameters.

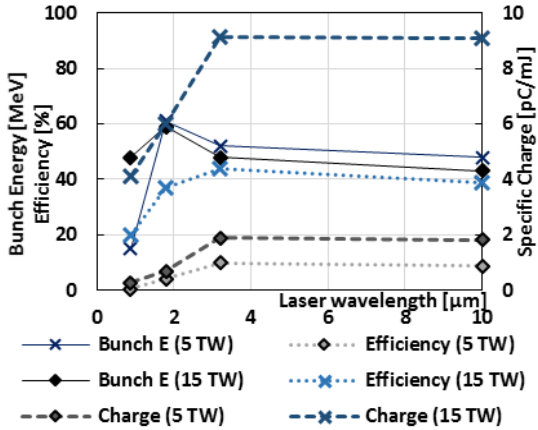

Fig. 3. The dependence of the specific bunch charge (charge/pulse energy ratio) (crosses) and the average bunch energy (squares) for the 5 TW (orange) and 15 TW (black) cases

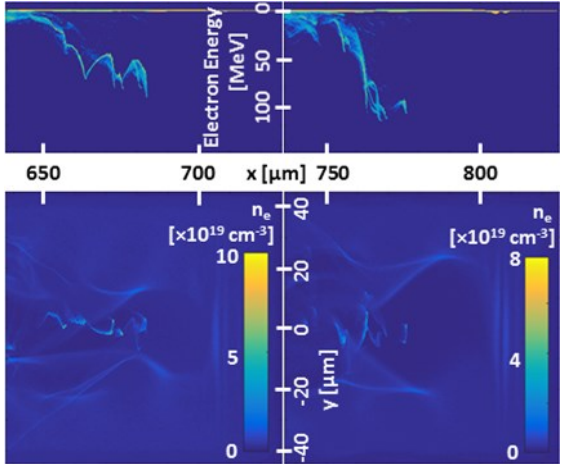

Fig: 4. x-$p_x$ phase-space plot (top, logarithmic, arb. u.) and 2D electron density map (bottom) for the 15 TW, 3.2 μm, 0.02$n_c$ case (left), and for the 0.015 $n_c$ (right).

All four instigated wavelengths behave similarly in the 15 TW case. There is continuous injection of electrons into the plasma bubble, with a corresponding, broad energy spectrum. This results in a high efficiency (upto 44%) electron injection and aceleration, with very high bunch charges, although the beam quality ends up very poor due to electrons being injected until the plasma bubble is completely disrupted.

The injection is dominantly density dependent, at half the electron density (0.01 $n_c$) no first-bubble injection was observed at the investigated parameters (only in later bubbles), even at 15 TW laser power, at any of the investigated wavelengths. One investigated intermediate-density case ($n_e$ = 0.015 $n_c$, 15 TW, 3.2 μm) showed similar behavior to the corresponding 0.02 $n_c$ case (Fig. 4.), i.e. continuous injection at somewhat higher energies - as expected for the lower density – although at lower bunch charge.

Furthermore, in the 10 μm/5 TW case only the second ionization electrons ('He II') are injected in the first bubble, in the rest of the cases (except 880 nm/5 TW both electrons from the first and second ionization (He I and He II) are injected in the first bubble.

## IV. Discussion

While the 5 TW/800 nm case showed only marginal injection, the 5 TW mid-IR cases showed quasi-monoenergetic features. The 15 TW cases all showed very high laser-to-electron energy conversion efficiency and inferred bunch charges, the continuous injection present results in low beam quality, and continuous electron spectra. This beam quality is expected to be improved by fine-tuning the simulated laser and plasma parameters at the expense of the bunch charge.

The high bunch charge is not specific to the EPOCH code, as it was also reproduced in simulations using the VORPAL code [21] for the 15 TW, 3.2 μm (highest efficiency) case, using similar simulation parameters.

The acceleration distance was approximately the depletion length $l_{pd} = c\tau_0 n_c/n_e$ [9], which, at the investigated laser and plasma parameters, was the same as the vacuum Rayleigh length of the prescribed laser pulse (within 7%).

Due to the nondimensional scaling, the self-injection criterion [22]

$$\frac{\alpha P}{P_c} > \frac{1}{16}\left[\ln\left(\frac{2n_c}{3n_e}\right) - 1\right]^3$$

is identical for the different wavelength cases (where α is the fraction of the laser energy within FWHM diameter and $P_c$ is the critical laser power for self focusing [23]), and is satified for α>0.2 in the 5 TW case. Identifying the root cause for the strong wavelength-dependence of the injection, i.e. the marginal injection at 880nm, would require further investigation.

The bunch energy decrease by the increasing wavelength could be attributed to beam loading effects, the highest average bunch energies were for 1.8 μm wavelengths, where the specific bunch charge was 35-70% of that in the 3.2 μm and 10 μm cases. The higher average bunch energy for the 5 TW case is due to the distinct electron bunch, while in the 15 TW cases such a bunch could not be distinguished due to the continuous injection. The maximum electron energies are higher for the 15 TW cases.

## V. Conclusions

The conducted 2D PIC parametric study of the wavelength effect on the LWFA process indicated that both efficiency and bunch charge (per laser pulse energy) is improved by increasing the wavelength of the driving laser, compared to the baseline λ=880 nm case. Simulation showed very high specific bunch charge, over 9 pC for every mJ of laser energy, and a correspondingly high – up to 44% - energy conversion efficiency for mid-IR (1.8-10 μm) wavelength laser drivers.

These beams are well suited for prospective use in X-ray generation, either via the inverse Compton scattering mechanism or via Bremsstrahlung in high Z slabs, as the very high bunch charge can allow for energy selectivity either in the electrons or generated X-rays, while maintaining a sufficient number of photons per selected energy bin.

**Acknowledgement**
The authors thank Zsolt Lécz for running the VORPAL comparison. The ELI-ALPS project (GINOP-2.3.6-15-2015-00001) is supported by the European Union and co-financed by the European Regional Development Fund.